\documentclass[superscriptaddress,reprint,aps,pra,showpacs]{revtex4-1}

\usepackage{amsmath, amssymb, graphicx, psfrag}

\usepackage{txfonts}

\def\Hz{\mbox{ Hz}}
\def\kHz{\mbox{ kHz}}

\def\G{\mbox{ G}}

\def\cm{\mbox{ cm}}

\def\mps{\mbox{ m/s}}

\newcommand{\ket}[1]{\left|#1\right>}

\begin{document}

\title{Time dependent spin-dressing using a $^3$He atomic beam}

\author{S. Eckel}
\author{S. K. Lamoreaux}
\affiliation{Yale University, Department of Physics, P.O. Box 208120, New Haven, CT 06520-8120}
\author{M. E. Hayden}
\affiliation{Simon Fraser University, Department of Physics, 8888 University Drive, Burnaby BC Canada V5A 1S6}
\author{T. M. Ito}
\affiliation{Physics Division, Los Alamos National Laboratory, Los Alamos, New Mexico 87545}

\pacs{11.30.Er, 13.40.Em, 32.10.Dk, 67.30.ep}

\date{\today}

\begin{abstract}
We have performed high precision experimental measurements of spin precession using a dressed $^3$He atomic beam.  Spin dressing uses an oscillating magnetic field that is both detuned to a high frequency and orthogonal to a static magnetic field to effectively change the gyromagnetic ratio of a spin.  We verify the validity of the spin-dressing Hamiltonian in regions beyond the limiting solution in which the Larmor frequency is much smaller than the frequency of the dressing field.  We also evaluate the effect of magnetic field misalignment, i.e. if the oscillating magnetic field is not orthogonal to the static magnetic field.  Modulation of the dressing field parameters is also discussed, with a focus on whether such a modulation can be approximated merely as a time dependent, dressed gyromagnetic ratio.  Furthermore, we discuss implications for a proposed search for the neutron electric dipole moment, which would employ spin-dressing to make the effective $^3$He and neutron magnetic moments equal.

\end{abstract}

\maketitle

\section{Introduction}

The existence of a permanent electric dipole moment (EDM) of a particle would be an indication of both parity (P) and time reversal (T) asymmetries in the fundamental interactions that describe the particle.  Because of the CPT theorem, which states that any Lorentz-invariant theory must conserve the combined symmetry operations of charge conjugation (C), parity, and time reversal, the existence of an EDM would be indicative of CP asymmetry as well~\cite{KhriplovichLamoreaux1997}.  Although cosmological evidence suggests that the fundamental laws of physics violate CP symmetry, the standard model of particle physics currently does not produce enough CP violation to explain the observed matter anti-matter asymmetry in the universe~\cite{Farrar1994}.  Extensions to the standard model, such as supersymmetry, tend to contain more CP violation and therefore produce larger EDMs~\cite{KhriplovichLamoreaux1997}.  Such large EDMs could be detected in the next generation of experiments.  Moreover, experimental searches for the neutron EDM ($n$EDM) place the most stringent limits on the so-called $\theta$ term in QCD.  The lack of this term in the standard model Lagrangian has led to the proposal of the axion~\cite{Peccei1977,Weinberg1978,Wilczek1978}.

A proposed experiment to detect the $n$EDM would use spin polarized $^3$He as a comagnetometer~\cite{Golub1994}.  When a neutron is absorbed by a $^3$He atom, the ensuing nuclear reaction releases 764 keV of energy, distributed as kinetic energy among the charged reaction products (a proton and a triton). If the reaction occurs in liquid helium, some of the kinetic energy lost to the liquid  produces scintillation light~\cite{Roberts1973,Surko1970}.  Because the absorption cross section depends on the relative orientation of the two spins~\cite{Passell1966}, i.e. $\mathbf{S}_n\cdot\mathbf{S}_3$, the scintillation signal in a bath of polarized $^3$He and neutrons precessing in a magnetic field would be time-modulated as $\cos [(\gamma_n-\gamma_3)Bt+\phi]$, where $\gamma_n$ and $\gamma_3$ are the gyromagnetic ratios of the neutron and $^3$He, respectively, and $\mathbf{S}_n$ and $\mathbf{S}_3$ are the spin vectors of the neutron and $^3$He, respectively.  The EDM of the $^3$He atom is expected, on firm theoretical grounds, to be vastly smaller than the $n$EDM.  The presence of an EDM would modify the precession of the neutrons if an electric field is applied and therefore change the modulation frequency of the scintillation signal.

\begin{figure}
 %
 %
 \providecommand\matlabtextA{\selectfont\strut}%
 \psfrag{027}[cl][cl]{\matlabtextA $\left|\gamma'B_0/\omega_d\right|$}%
 \psfrag{028}[bc][bc]{\matlabtextA $E/\hbar\omega_d - \left<n\right>$}%
 \psfrag{029}[tc][tc]{\matlabtextA $x = \gamma B_d/\omega_d$}%
 %
 %
 %
 \def\matlabfragNegXTick{\mathord{\makebox[0pt][r]{$-$}}}
 \providecommand\matlabtextB{\tiny\selectfont\strut}%
 \psfrag{000}[ct][ct]{\matlabtextB $2.2$}%
 \psfrag{001}[ct][ct]{\matlabtextB $2.3$}%
 \psfrag{002}[ct][ct]{\matlabtextB $2.4$}%
 \providecommand\matlabtextC{\footnotesize\selectfont\strut}%
 \psfrag{008}[ct][ct]{\matlabtextC $0$}%
 \psfrag{009}[ct][ct]{\matlabtextC $1$}%
 \psfrag{010}[ct][ct]{\matlabtextC $2$}%
 \psfrag{011}[ct][ct]{\matlabtextC $3$}%
 \psfrag{012}[ct][ct]{\matlabtextC $4$}%
 \psfrag{013}[ct][ct]{\matlabtextC $5$}%
 %
 %
 %
 \providecommand\matlabtextD{\tiny\selectfont\strut}%
 \psfrag{003}[rc][rc]{\matlabtextD $-0.02$}%
 \psfrag{004}[rc][rc]{\matlabtextD $-0.01$}%
 \psfrag{005}[rc][rc]{\matlabtextD $0$}%
 \psfrag{006}[rc][rc]{\matlabtextD $0.01$}%
 \psfrag{007}[rc][rc]{\matlabtextD $0.02$}%
 \providecommand\matlabtextE{\footnotesize\selectfont\strut}%
 \psfrag{014}[rc][rc]{\matlabtextE $-1$}%
 \psfrag{015}[rc][rc]{\matlabtextE $-0.8$}%
 \psfrag{016}[rc][rc]{\matlabtextE $-0.6$}%
 \psfrag{017}[rc][rc]{\matlabtextE $-0.4$}%
 \psfrag{018}[rc][rc]{\matlabtextE $-0.2$}%
 \psfrag{019}[rc][rc]{\matlabtextE $0$}%
 \psfrag{020}[rc][rc]{\matlabtextE $0.2$}%
 \psfrag{021}[rc][rc]{\matlabtextE $0.4$}%
 \psfrag{022}[rc][rc]{\matlabtextE $0.6$}%
 \psfrag{023}[rc][rc]{\matlabtextE $0.8$}%
 \psfrag{024}[rc][rc]{\matlabtextE $1$}%
 %
 \includegraphics{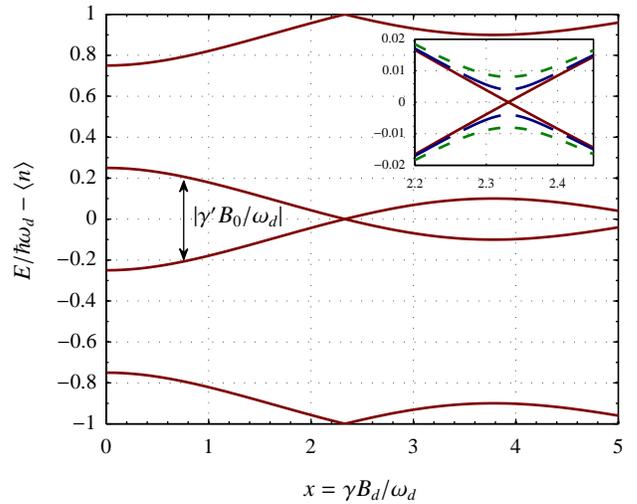}
 \caption{\label{fig:states} (Color online) Eigenvalues of the spin-dressing Hamiltonian (Eq.~\ref{eq:ham2}) as a function of dressing parameter $x$ in units of the energy of the dressing field photons for $y=0.5$.  The difference between the two central states is associated with the dressed gyromagnetic ratio $\gamma'$ for that value of $x$ and $y$ (see text for full explanation).  (Inset) Appearance of an avoided crossing near $x=2.32$ when the angle between the dressing field $B_d$ and static field $B_0$ is $90^\circ$ (solid line), $89^\circ$ (dashed line) and $88^\circ$ (short dashed line).}
\end{figure}

The magnetic moments of the neutron and $^3$He atom are equal to within 12\% ($\gamma_n/2\pi=-2.916\mbox{ kHz/G}$~\cite{CODATA2006} and $\gamma_3/2\pi=-3.243\mbox{ kHz/G}$~\cite{Williams1969,Flowers1993}, respectively), and so monitoring the differential precession frequency as described above reduces by nearly an order of magnitude the undesired effects of external magnetic field noise and systematic variations of magnetic fields~\cite{Golub1994}.  An ideal experiment would compare a neutron to an atom that has exactly the same magnetic moment.  No such atom exists, and furthermore the only atom that will remain in solution with liquid helium at low temperatures is $^3$He.  However, it is possible to tune the effective magnetic moment of a particle by the use of spin-dressing, and, moreover, to make the effective  gyromagnetic ratios of the neutron and $^3$He atom equal.  This procedure is referred to as critical dressing. The parameters to achieve critical dressing depend only weakly (to the second order) on the static magnetic field.

An added benefit of spin dressing is that the $^3$He and neutron spins can be held closer to parallel on average, thereby reducing the rate of absorption of neutrons by the $^3$He magnetometer atoms.  As discussed in Ref.~\cite{Golub1994}, by modulating the dressing field amplitude about the point where $\langle\hat{S}_n\cdot\hat{S}_3\rangle = 1$, the scintillation light will be modulated at twice the dressing field modulation frequency.  If $\langle\hat{S}_n\cdot\hat{S}_3\rangle \neq 1$, the scintillation light will have a frequency component at the modulation frequency.  Feedback can then be employed on the DC component of the dressing field amplitude to drive the modulation frequency harmonic of the scintillation light to zero which will hold the neutron spins parallel to the $^3$He spins on average.  This increases the effective coherence time while maintaining the same average signal amplitude (scintillation rate) compared to simple Larmor precession in a static magnetic field.  

\begin{figure}
 %
 %
 \providecommand\matlabtextA{\selectfont\strut}%
 \psfrag{016}[cl][cl]{\matlabtextA Increasing $y$}%
 \psfrag{017}[bc][bc]{\matlabtextA $\gamma '/\gamma$}%
 \psfrag{018}[cc][cc]{\matlabtextA $x = \gamma B_d/\omega_d$}%
 %
 %
 %
 \def\matlabfragNegXTick{\mathord{\makebox[0pt][r]{$-$}}}
 \providecommand\matlabtextB{\footnotesize\selectfont\strut}%
 \psfrag{000}[ct][ct]{\matlabtextB $0$}%
 \psfrag{001}[ct][ct]{\matlabtextB $1$}%
 \psfrag{002}[ct][ct]{\matlabtextB $2$}%
 \psfrag{003}[ct][ct]{\matlabtextB $3$}%
 \psfrag{004}[ct][ct]{\matlabtextB $4$}%
 \psfrag{005}[ct][ct]{\matlabtextB $5$}%
 \psfrag{006}[ct][ct]{\matlabtextB $6$}%
 \psfrag{007}[ct][ct]{\matlabtextB $7$}%
 %
 %
 %
 \psfrag{008}[rc][rc]{\matlabtextB $-0.4$}%
 \psfrag{009}[rc][rc]{\matlabtextB $-0.2$}%
 \psfrag{010}[rc][rc]{\matlabtextB $0$}%
 \psfrag{011}[rc][rc]{\matlabtextB $0.2$}%
 \psfrag{012}[rc][rc]{\matlabtextB $0.4$}%
 \psfrag{013}[rc][rc]{\matlabtextB $0.6$}%
 \psfrag{014}[rc][rc]{\matlabtextB $0.8$}%
 \psfrag{015}[rc][rc]{\matlabtextB $1$}%
 %
 \includegraphics{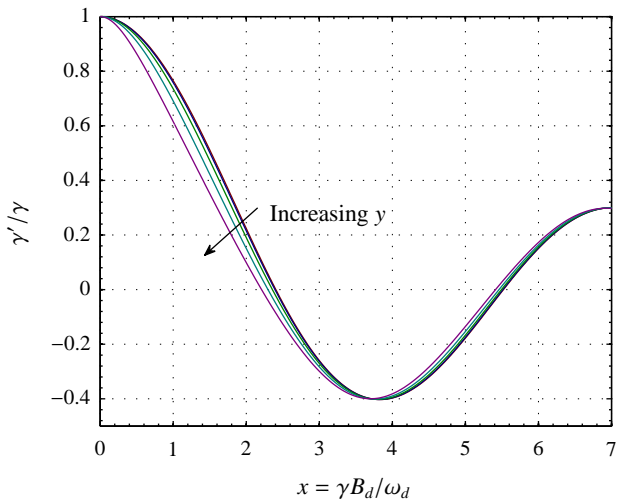}
 \caption{\label{fig:effective_gamma} (Color online) Effective gyromagnetic ratio as a function of the dressing field amplitude, $x$.  Each curve represents a different value of $y$; the values of $y$ that are plotted are $0$, $0.2$, $0.4$, $0.6$, and $0.8$.  The arrow indicates the direction of increasing $y$.  The curve for $y=0$ is merely $J_0(x)$, which is the analytic solution to the spin-dressing Hamiltonian.}
\end{figure}
 
This proposed implementation of the dressed spin technique requires that the dressing parameters be modulated to infer the critical dressing condition; a $n$EDM would be evident by a change in the critical dressing condition as a function of the electric-field direction relative to the applied magnetic field.  To date, no high-accuracy measurements or full calculations of the influence of relatively slow modulation on the critical dressing parameters have been performed. Furthermore, the theory for the effects of static field magnitude and misalignment, presented in Ref.~\cite{Golub1994} as perturbative effects, have not been verified.  The goals of the work presented here are to experimentally verify the theory presented in Ref.~\cite{Golub1994} at a level of accuracy suitable for the $n$EDM experiment and to extend the details of the calculations to encompass the possible effects of modulation of the dressing parameters.

Spin dressing uses an oscillating magnetic field detuned to high frequency (the dressing field) to effectively change the gyromagnetic ratio of a spin-$^1/_2$ particle~\cite{Cohen-Tannoudji1969}.  The Hamiltonian for this system can be written as
\begin{equation}
 \label{eq:ham1}
 H=-\gamma B_0 \hat{s}_z + \hbar \omega_d \hat{a}^\dagger\hat{a}+\frac{\gamma B_d}{2\sqrt{\langle n \rangle}}\hat{s}_x(\hat{a}+\hat{a}^\dagger)\ ,
\end{equation}
where $\langle n \rangle\gg1$ represents the average number of photons in the coherent, oscillating spin-dressing field, $B_d$ is the amplitude of the dressing magnetic field, $\omega_d$ is the angular frequency, $B_0$ is the static magnetic field along the $\hat{z}$ direction, and $\gamma$ is the undressed gyromagnetic ratio for the spin of interest.  The operators $\hat{s}_i$ and $\hat{a}$ correspond to the spin angular momentum and photon annihilation operators, respectively.  This Hamiltonian can be rewritten by defining the quantities $y\equiv \gamma B_0/\omega_d$ and $x\equiv\gamma B_d/\omega_d$ as in Ref.~\cite{Esler2007}, making Eq.~\ref{eq:ham1} become
\begin{eqnarray}
 \frac{H}{\hbar \omega_d} - \hat{a}^\dagger\hat{a}& = & -\frac{y}{2} \hat{\sigma}_z + \frac{x}{4\sqrt{\langle n \rangle}}(\hat{\sigma}_-\hat{a} +\hat{\sigma}_-\hat{a}^\dagger \nonumber \\
  & & \ \ \ \ \ \ +\hat{\sigma}_+\hat{a}+\hat{\sigma}_+\hat{a}^\dagger) \label{eq:ham2}\ ,
\end{eqnarray}
where $\hat{\sigma}_i$ are the usual Pauli spin matrices.  The eigenvalues of this Hamiltonian, when diagonalized in the $\left|n,\pm\right>$ basis, are shown in Fig.~\ref{fig:states}.  The difference in energy between the two eigenstates around $E=0$ is then $|\hbar\gamma'B_0|$, where $\gamma'$ is the dressed gyromagnetic ratio of the spin, which depends on $x$ and $y$.  The relative sign of $\gamma'$ is determined by both the initial sign of $\gamma$ and by diabatically following the states as $x$ is increased.  In the limit of $y\ll1$, the solution yields $\gamma' = \gamma J_0(x)$, where $\gamma'$ and $\gamma$ are the dressed and undressed gyromagnetic ratios, respectively, and $J_0$ is the zeroth order Bessel function~\cite{Cohen-Tannoudji1969}.  The ratio of the dressed to undressed gyromagnetic ratios, $\gamma'/\gamma$, is shown in Fig.~\ref{fig:effective_gamma} as a function of $x$ for multiple values of $y$.  Because of the structure of the Hamiltonian, the ratio of $\gamma'/\gamma$ is identical for $\gamma>0$ and $\gamma<0$.

In the specific case of the proposed $n$EDM experiment, we are interested in the amplitude of the dressing field that yields the critical dressing, at which point the two dressed gyromagnetic ratios are equal to each other.  For $y\ll1$, this occurs when $\gamma_3'=\gamma_n'$ or $\gamma_3 J_0(x_3) = \gamma_n J_0(x_n)$, where $x_i = \gamma_i B_d/\omega_d$.  Because $\gamma_3\approx1.11\gamma_n$, the point at which the dressed gyromagnetic ratios are equal is when $|x_3| \approx1.31$  (or equivalently, $|x_n|\approx 1.18$) in the limit when $y\ll1$.

In this paper, we experimentally verify the spin-dressing predictions described above and in Ref.~\cite{Golub1994}, with particular focus on the critical dressing point.  There have been two previous measurements of spin dressing in $^3$He, one in a cell~\cite{Chu2011} and one in a beam experiment~\cite{Esler2007}.  The latter is similar to this work and employed many of the same components.  However, several important changes were made that brought an improvement in accuracy and allow for a detailed comparison between theory and experiment at a level relevant for the proposed $n$EDM experiment described in Ref.~\cite{Golub1994}.  The apparatus and the improvements are detailed in Sec.~\ref{sec:apparatus}.  By numerically solving the Hamiltonian (Eq.~\ref{eq:ham2}) for conditions appropriate to our experiment and comparing the results with our data,
we verify the perturbation calculations presented in Ref.~\cite{Golub1994}.  These results are presented in Sec.~\ref{sec:results}.  We also search for effects that would prevent one from simply approximating the effect of modulating the amplitude of the dressing field as being equivalent to quasi-statically changing the spin-precession frequency.  With our current level of precision, such effects were not conclusively seen.  Our inability to experimentally detect such effects in a system where the dressing parameters are varied rapidly are supported by numerical simulations and theoretical analysis.  These calculations are discussed in Sec.~\ref{sec:adiabatic}.

Because this work will focus exclusively on $^3$He, we will henceforth use $\gamma$ and $\gamma'$ to refer to the dressed and undressed gyromagnetic ratios of the nuclear spin of $^3$He.  However, this spin-dressing technique can, in principle, be applied to any spin-$^1/_2$ particle.

\section{Apparatus}
\label{sec:apparatus}

\begin{figure*}
 \center
 \psfrag{001}[tc][tc]{$\pi/2$ coil}
 \psfrag{002}[bc][bc]{Polarized $^3$He}
 \psfrag{003}[bl][bl]{Solenoid ($l_s=90$~cm)}
 \psfrag{004}[bl][bl]{Ferromagnetic shield ($l_\text{shield}=83$~cm)}
 \psfrag{005}[c][c]{$L=51$~cm}
 \psfrag{006}[c][c]{Compensating coils}
 \psfrag{007}[c][c]{$l_d = 18$~cm}
 \psfrag{008}[c][c]{Compensating field}
 \psfrag{009}[tl][tl]{Dressing field}
 \psfrag{010}[c][c]{Dressing coil}
 \psfrag{011}[c][c]{Solenoid field}
 \psfrag{012}[c][c]{Compensating field}
 \psfrag{013}[c][c]{$l_c = 1.2$~cm}
 \psfrag{014}[tc][tc]{$\pi/2$ coil}
 \psfrag{015}[c][c]{To analyzer}
 \psfrag{016}[r][r]{$l_{sf}$}
 \psfrag{017}[l][l]{$l_{sf}=3.2\cm$}
 \includegraphics{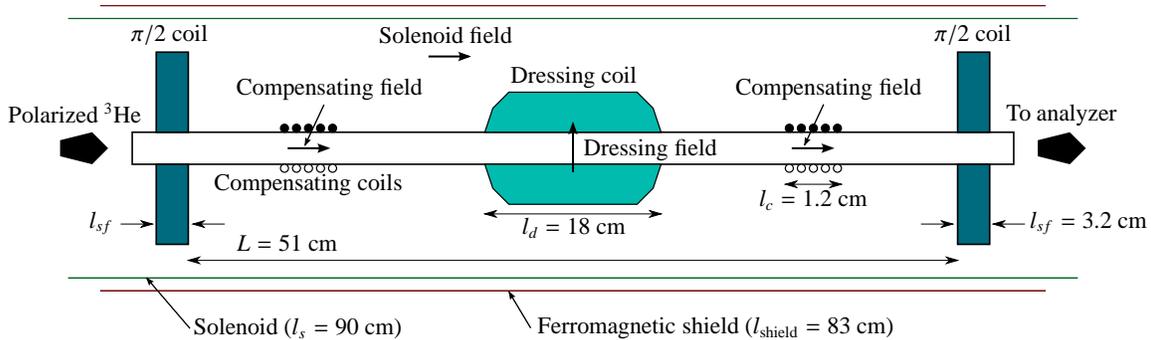}
 \caption{\label{fig:schematic} (Color online) A schematic drawing of the apparatus used to measure the $^3$He spin precession (see text for description).  Not shown are the atomic beam source, the polarizing magnet, the analyzer magnet, and the RGA used to prepare and detect the $^3$He.  The coordinate system has the beam propagating in the $\hat{z}$ direction, the $\hat{x}$ direction pointing up, and the $\hat{y}$ direction pointing out of the page.}
\end{figure*}

We investigate spin precession using Ramsey's method of separated oscillatory fields (SOF).  A schematic drawing of the apparatus is shown in Fig.~\ref{fig:schematic}.  In the main experimental region, a large solenoid produces a homogeneous, static magnetic field in the $+\hat{z}$ direction.  In addition to this solenoid, there are five separate magnetic field coils: two $\pi$/2 rotation (or ``spin flip'') coils, two ``compensating field'' coils, and a ``dressing field'' coil.  The role played by each coil is described briefly below, and the construction and operation of each coil is detailed in the following several subsections.  These coils are contained within a ferromagnetic shield to minimize the effect of external magnetic fields.  

A beam of spin-polarized $^3$He atoms from a source (discussed below) enters the experimental region from the left in Fig.~\ref{fig:schematic}.  The first and last coils encountered by the $^3$He are the spin flip coils, which are used to apply oscillating magnetic fields that induce $\pi/2$ spin rotations.  The first $\pi/2$ coil tips the $^3$He spins from $+\hat{z}$ to the $\hat{x}$-$\hat{y}$ plane.  The spins then precess about the static magnetic field as they travel to the second $\pi/2$ coil where the spin component perpendicular to the appropriate rotating component of the oscillating field is flipped into the $-\hat{z}$ direction.  The detector system, which consists of an analyzer magnet and residual gas analyzer (discussed below), detects only those $^3$He atoms with their spins aligned along $+\hat{z}$.  The phase of the oscillating field in the two coils is the same; therefore the transmitted intensity is proportional to $\frac{1}{2}\left[1-\cos(\phi-\phi_{sf})\right]$, where $\phi$ is the total phase accumulated by the spins as they precess about the $+\hat{z}$ axis in between the two $\pi/2$ coils, and $\phi_{sf}$ is the phase accumulated by the spin-flip field during that time.  The minus sign stems from the fact that we detect $^3$He with spin projection along $+\hat{z}$.  Because the spin flip fields are oscillating and linearly polarized (as opposed to rotating or circularly polarized), our apparatus is completely insensitive to the sign of the gyromagnetic ratio.  For this reason, we will take $\gamma>0$ for clarity and note that the results of this experiment are independent of the sign of the gyromagnetic ratio.

As can be seen from the above discussion, the minimum transmission occurs when the phase of the spin-flip field ($\phi_{sf}$) equals the phase accumulated by the spins ($\phi$) during their propagation between the $\pi/2$ coils.  The phase accumulated by the spin flip field is given by $\phi_{sf} = \omega L/v$, where $\omega$ is the spin-flip field frequency, $L$ is the distance between the two spin-flip coils, and $v$ is the velocity (of a particular group of atoms in the beam).  The phase accumulated by the spins is given by
\begin{eqnarray}
 \phi & = & \int_{-L/2}^{L/2} \gamma'(z) B(z)\ \frac{dz}{v} + \frac{2}{\pi}\left[\int_{-(L/2+l_{sf})}^{-L/2} \gamma \left(B(z)-\frac{\omega}{\gamma}\right)\ \frac{dz}{v} \right. \nonumber \\ 
 & & \label{eq:priint} \left. + \int_{L/2}^{L/2+l_{sf}} \gamma \left(B(z)-\frac{\omega}{\gamma}\right)\ \frac{dz}{v} \right]\ ,
\end{eqnarray}
where $l_{sf}$ is the effective length of the oscillating magnetic field generated by each of the spin flip coils and $B(z)$ is the value of the static magnetic field at position $z$.  Here, $z=0$ is the geometric center of the experiment.  The first integral in Eq.~\ref{eq:priint} represents precession of the spins in between the $\pi/2$ spin flip coils and the last two integrals in Eq.~\ref{eq:priint} represent the detuning of the static field from the resonance condition ($\omega = \omega_0 = \gamma B$) while the spins are located within the spin flip coils.  

Centered in the experiment is the dressing coil, which applies an oscillating magnetic field in the $+\hat{x}$ direction that serves as the dressing field.  Because $\gamma'<\gamma$, the principal effect of the dressing field is to slow the spin precession down and therefore reduce $\phi$.  Because the dressing field only extends over a small length ($\sim18\cm$) compared to the distance between the $\pi/2$ spin-flip coils ($\sim 51\cm$), the value of $\gamma'$ is dependent on $z$ and therefore appears as a function of $z$ in Eq.~\ref{eq:priint}.  Midway between the dressing coil and the $\pi/2$ coils are two compensating coils that apply an additional, static field in the $+\hat{z}$ direction.  This additional static field makes the spins precess faster while in these coils and therefore increases $\phi$.  One may then adjust the compensating field strength for a given dressing field so that the total phase difference between the spins and the spin flip field satisfies $\Delta\phi=\phi - \phi_{sf} = 0$.  As seen by equating $\phi_{sf} = \omega L/v$ with the expression for $\phi$ (Eq.~\ref{eq:priint}), this condition is achieved independent of $v$.  If $\Delta\phi\neq 0$, then $\cos^2\Delta \phi$ is not unity and interpreting the transmitted intensity as a frequency shift requires detailed knowledge of the beam velocity spectrum.  This problem is avoided in our measurements by determining the compensating coil current that forces $\Delta \phi=0$.

The effects of varying the dressing field strength are determined as follows.  First, with no DC current applied to the compensating coils and no AC current applied to the dressing coil, the static field is tuned such that $\Delta\phi=0$ for a fixed $\omega$.  Then, with the dressing field frequency set to a specific value (for a set of measurements), the dressing field amplitude is increased in discrete steps. At each step, a current is applied to the compensating coils to bring the $^3$He transmission back to a minimum.  Thus the $\pi/2$ coils are operated continuously at their proper frequency for a given static field, unlike Ref~\cite{Esler2007} where the spin-flip field's frequency is varied in an attempt to determine the frequency shift due to the dressing effect.

\subsection{Atomic Beam Source and Polarizer}

The $^3$He atomic beam is produced by a 1.2~cm diameter tube packed with 1~mm capillaries, each 3.5~cm long.  The nozzle is held at 1.4~K by use of a pumped liquid helium cell.  $^3$He gas is delivered to the nozzle through a cooled 15~cm long capillary, which is in turn connected to a sequentially-cooled 2~mm ID tube, supplied with $^3$He at room temperature with pressure 1.5~torr.  The beam is directed through a 1~m long quadrupole spin state selecting magnet (polarizer) constructed from NdFeB permanent magnets (field at pole surface of 7.5~kG), with an effective aperture of 1.5~cm.  The polarized atom flow rate at the output can exceed 10$^{14}$ atoms/sec.

The polarized $^3$He beam is transported through the apparatus inside a ``standard length'', 48" long, 32 mm diameter, Pyrex tube.  The dressing coil and the two compensating coils are mounted directly onto this Pyrex tube.  Inside this portion of the apparatus, the polarization achieved is approximately 95\%.  The fact that it is less than 100\% is due to the $^3$He background pressure, which is in turn a result of the inefficiency in pumping $^3$He.

\subsection{Solenoid, Magnetic Shield, and the $\pi/2$ coils}
\label{sec:piby2coils}
The homogeneous static magnetic field is produced by a solenoid that is surrounded by a 1~mm thick, Co-Netic AA stress annealed shield.  The solenoid has 780~turns wound on a mean diameter of 15.36~cm and is 90.5~cm long.  The shield has a length of 83~cm and has an inner diameter of 26.5~cm.

The two $\pi/2$ coils were identically constructed and supported by the G-10 glass epoxy tube onto which the solenoid was wound.  Each is wound using 1~mm diameter wire wrapped on a plywood form in the shape of a 1.9~cm by 7.6~cm rectangular solenoid. The coil is split axially into two identical series driven sections separated by 32~mm, each with 40 turns. The axis of this coil is mounted so as to produce a field directed perpendicular to the beam, with the short (1.9~cm) dimension of the coil being aligned parallel to the beam.  The split coil is then sandwiched between two 15~cm diameter disks of 0.4~mm thick Cu. A 32~mm diameter hole is cut in the center of each disk to allow the beam tube to pass through. To facilitate mounting around the beam tube, the Cu disks are each cut in half along a diameter and then held together with Cu tape. The cuts are orientated parallel to the windings; eddy current flow is largely parallel to the cut and the shielding is not affected.  Mapping the oscillating field with a small pickup coil shows that the field maximum at the center of the coil, on the beam axis, falls by a factor of two at the Cu sheet, and at farther distances falls exponentially with characteristic length of order of the hole radius.  Sensitive measurements with the pickup loop showed insignificant leakage field from the $\pi/2$ coils in all regions external to the Cu shielding disks. The two coils are spaced by 51~cm.

The coils are connected in series, which ensures equal current in both with no relative phase shift.  In addition, a series capacitor (determined by chosen frequency) is used to cancel the reactive component of the coils' impedance.  The current is measured using a series connected 10~$\Omega$, 2~W resistor.  The current is supplied by a National Semiconductor LM4700 30~Watt audio amplifier IC, which is driven with a sinusoidal voltage from an Agilent 34401 arbitrary function generator. The spin-flip field amplitude was adjusted to give a minimum transmission on resonance.  Note that the transmitted $^3$He beam intensity as a function of the spin-flip field amplitude on resonance provides a measure of the Fourier power spectrum of the beam velocity distribution.  This spectrum can be transformed to the velocity probability distribution which has a peak at approximately 140~m/s. The $\pi/2$ condition typically required a current of a few hundred mA.

\subsection{Analyzer and Detector}

A second state selecting magnet identical to the polarizer is used as a spin analyzer.  The input aperture of the analyzer is located approximately 180 cm from the output aperture of the polarizer.  The Pyrex tube connects the polarizer and analyzer vacuum systems which are otherwise separate.  Guide fields from the polarizer (analyzer) output (input) are produced by two single pancake coils, each 45~cm in diameter, located midway between the output (input) apertures and the magnetic shield.    

The detector is a Stanford Research Systems Residual Gas Analyzer (RGA) model 100 with a channel electron multiplier.  The RGA is mounted on the side port of a conflat tee section, with the RGA axis perpendicular to the beam axis.  The beam enters the tee through a 15 cm-long, 1.5 cm ID tube and strikes a blank flange mounted on the opposing port. In effect, the tee acts as a pressure buildup volume, allowing the $^3$He pressure to increase by nearly a factor of 20 relative to that which is obtained when the beam is delivered directly to the RGA.  The RGA is set to mass 3, and the pressure is read from the RS-232 port at a maximum rate of about 8 Hz by use of a simple MS QUICK BASIC program.  The signal-to-noise ratio is about 100 per reading, with a background signal corresponding to unpolarized $^3$He or gas which reads as mass 3.

In order to determine the compensating coil current required to give the minimum $^3$He transmission, the drive frequency for the $\pi/2$ coils is modulated at 2~Hz with a deviation of 100~Hz.  This is slightly smaller than the expected linewidth for the SOF method which is given by
\begin{equation}
 \delta f\ [{\rm Hz}]=\frac{1}{2T}= \frac{v}{2L}=\frac{140\ {\rm m/s}}{2\times0.51\ {\rm m}} \approx 140\mbox{ Hz}.
\end{equation}
The digital reading from the RGA is converted to a voltage with an 8~bit digital to analog converter and is phase-locked to the 2~Hz signal using a lock-in amplifier.  The compensating coil current can then be readily set to null the detected signal at the modulation frequency.  In the vicinity of this minimum, the $^3$He transmission probability only depends on the phase difference to the second order [i.e., $\cos(\Delta\phi) \approx 1 - (\Delta\phi)^2$]. Therefore, when the signal is set to null at the modulation frequency, it is an indication that the average phase has been reset to zero.  For the measurements presented here, the adjustment of the dressing parameters and the setting of the compensating coil current and its measurement were done manually.

\subsection{Compensating Coils}

The two compensating coils are symmetrically located and wound directly on the beam tube and placed between the dressing coil and the two $\pi/2$ coils, approximately 6~cm away from the $\pi/2$ spin flip coils.  Each coil has 11 turns of 1~mm diameter wire and an effective diameter of 1.65~cm. Equal current is applied to the coils by connecting them in series, and their fields are orientated in the same direction.  A Fluke 87 True RMS Multimeter that was calibrated against an Agilent 34401A Digital Multimeter is used to measure the compensating current, which is the current required to return the overall phase to the minimum transmission condition.

The apparatus is sufficiently long that the field integral of the compensating coils, i.e.
\begin{equation}
\int_{-L/2}^{L/2} B_{c}(z)dz \approx n\mu_0 I    
\end{equation}
where $B_c$ is the compensating field as a function of $z$, $n$ is the total number of turns in the two coils, and $I$ is the current, is nearly independent of the coil radius and their position in the apparatus.  The long-distance field does penetrate into the spin flip and dressing coil regions and is taken into account in the analysis.  

\subsection{Dressing Cosine Coil}

A $\cos\phi$-like (or cylindrical saddle-shaped) coil~\cite{Bidinosti2005} placed midway between the two $\pi/2$ coils is used to apply the dressing field in the $\hat{x}$ direction. It was constructed using 1~mm diameter wire wound on a length of 3~in. (7.6~cm) ID PVC tube into which appropriate grooves had been machined on the outer surface. Each layer of the coil has nine axial current rungs per quadrant, which are then interconnected by a series of azimuthal current paths at either end (forming two spiral sets of saddle-shaped paths on the outer surface of the former).  A total of three complete windings are employed, resulting in a mean coil diameter of 8.5~cm. The set of azimuthal current paths at either end of the coil each span an axial distance of 1.6~cm and were positioned to give a mean length to diameter ratio of 2 and a total coil length of 18~cm.  As designed, the calculated central field produced by this coil is 4.3 G/A in the absence of shielding~\cite{Bidinosti2005,BioSavart}, and its measured inductance is 0.3~mH.

The cosine coil is driven in a fashion similar to the $\pi/2$ coils with the same model generator and amplifier.  However, it is driven with much higher currents, up to 4~A RMS maximum.  Again, a series capacitor is used to tune out the reactive component of the coil impedance.  The current is monitored using 8 parallel-connected 11 $\Omega$ 2 watt carbon composition resistors.  The resistance was calibrated as a function of current and frequency to check for possible nonlinear heating or other effects.  There is no frequency effect, however a small heating effect with increasing current was taken into account in the analysis.  An Agilent 34401A True RMS Multimeter is used to measure the voltage across the current monitor resistor for the dressing measurements and calibrations.  

Provisions were made to add a DC current in addition to the oscillating current in the dressing coil, allowing both the determination of the absolute average angle between the dressing field and static field, and to study the effects of intentional misalignment.  This was accomplished by use of a choke (inductor) of approximately 30~mH, connected to the node between the capacitor and dressing coil.  The choke presents a reactive impedance of 500 $\Omega$ at the lowest dressing frequencies,  much higher than the dressing coil impedance. Only the current through the dressing coil was sent through the monitor resistor.  There was no net effect on the AC current due to the presence of the choke.  

\section{Spin Dressing Results}
\label{sec:results}
\begin{figure*}
 %
 %
 \providecommand\matlabtextA{\selectfont\strut}%
 \psfrag{040}[cc][cc]{\matlabtextA $x = \gamma B_d/\omega_d$}%
 \psfrag{041}[bc][bc]{\matlabtextA Compensating field strength ($B_{c,av}/B_0$)}%
 \psfrag{042}[cc][cc]{\matlabtextA $x = \gamma B_d/\omega_d$}%
 \psfrag{043}[bc][bc]{\matlabtextA Compensating field strength ($B_{c,av}/B_0$)}%
 %
 %
 %
 \def\matlabfragNegXTick{\mathord{\makebox[0pt][r]{$-$}}}
 \providecommand\matlabtextB{\footnotesize\selectfont\strut}%
 \psfrag{000}[ct][ct]{\matlabtextB $0$}%
 \psfrag{001}[ct][ct]{\matlabtextB $0.5$}%
 \psfrag{002}[ct][ct]{\matlabtextB $1$}%
 \psfrag{003}[ct][ct]{\matlabtextB $1.5$}%
 \psfrag{004}[ct][ct]{\matlabtextB $2$}%
 \psfrag{005}[ct][ct]{\matlabtextB $2.5$}%
 \psfrag{006}[ct][ct]{\matlabtextB $3$}%
 \psfrag{007}[ct][ct]{\matlabtextB $3.5$}%
 \psfrag{008}[ct][ct]{\matlabtextB $4$}%
 \psfrag{009}[ct][ct]{\matlabtextB $4.5$}%
 \psfrag{020}[ct][ct]{\matlabtextB $0$}%
 \psfrag{021}[ct][ct]{\matlabtextB $1$}%
 \psfrag{022}[ct][ct]{\matlabtextB $2$}%
 \psfrag{023}[ct][ct]{\matlabtextB $3$}%
 \psfrag{024}[ct][ct]{\matlabtextB $4$}%
 \psfrag{025}[ct][ct]{\matlabtextB $5$}%
 \psfrag{026}[ct][ct]{\matlabtextB $6$}%
 \psfrag{027}[ct][ct]{\matlabtextB $7$}%
 \psfrag{028}[ct][ct]{\matlabtextB $8$}%
 \psfrag{029}[ct][ct]{\matlabtextB $9$}%
 %
 %
 %
 \psfrag{010}[rc][rc]{\matlabtextB $0$}%
 \psfrag{011}[rc][rc]{\matlabtextB $0.05$}%
 \psfrag{012}[rc][rc]{\matlabtextB $0.1$}%
 \psfrag{013}[rc][rc]{\matlabtextB $0.15$}%
 \psfrag{014}[rc][rc]{\matlabtextB $0.2$}%
 \psfrag{015}[rc][rc]{\matlabtextB $0.25$}%
 \psfrag{016}[rc][rc]{\matlabtextB $0.3$}%
 \psfrag{017}[rc][rc]{\matlabtextB $0.35$}%
 \psfrag{018}[rc][rc]{\matlabtextB $0.4$}%
 \psfrag{019}[rc][rc]{\matlabtextB $0.45$}%
 \psfrag{030}[rc][rc]{\matlabtextB $0$}%
 \psfrag{031}[rc][rc]{\matlabtextB $0.05$}%
 \psfrag{032}[rc][rc]{\matlabtextB $0.1$}%
 \psfrag{033}[rc][rc]{\matlabtextB $0.15$}%
 \psfrag{034}[rc][rc]{\matlabtextB $0.2$}%
 \psfrag{035}[rc][rc]{\matlabtextB $0.25$}%
 \psfrag{036}[rc][rc]{\matlabtextB $0.3$}%
 \psfrag{037}[rc][rc]{\matlabtextB $0.35$}%
 \psfrag{038}[rc][rc]{\matlabtextB $0.4$}%
 \psfrag{039}[rc][rc]{\matlabtextB $0.45$}%
 %
 \includegraphics{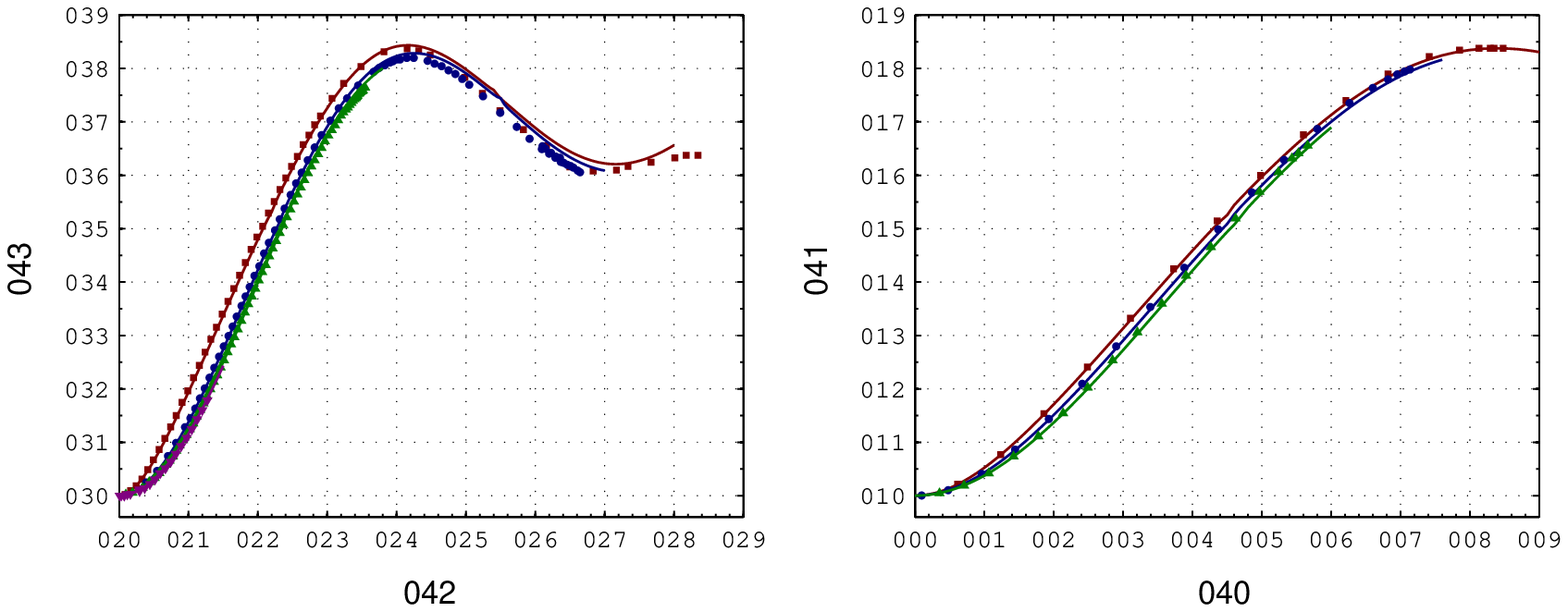}
 \caption{\label{fig:results} (Color online) Left: Average required (integrated) compensating field strength in units of the undressed magnetic field for an undressed precession frequency of 3.99~kHz and dressing field frequencies 5.12~kHz ($y = 0.779$, squares), 8.59~kHz ($y = 0.464$, circles), 16.4~kHz ($y = 0.243$, upward triangles), and 35.8 kHz ($y = 0.111$, downward triangles).  Right: Average required compensating field strength in units of the undressed magnetic field for an undressed precession frequency of 9.36~kHz and dressing field frequencies 13.75~kHz ($y = 0.681$, squares), 16.44~kHz ($y = 0.569$, circles), and 20.50~kHz ($y = 0.457$, upward triangles)}
\end{figure*}
\begin{figure}
 %
 %
 \providecommand\matlabtextA{\small\selectfont\strut}%
 \psfrag{028}[bl][bl]{\matlabtextA $\times\frac{1}{4}$}%
 \providecommand\matlabtextB{\selectfont\strut}%
 \psfrag{026}[tc][tc]{\matlabtextB Frequency (kHz)}%
 \psfrag{027}[bc][bc]{\matlabtextB Dressing coil field max (G/A)}%
 \psfrag{029}[bc][bc]{\matlabtextB Field strength (G/A)}%
 \psfrag{030}[tc][tc]{\matlabtextB Position from center (cm)}%
 %
 %
 %
 \def\matlabfragNegXTick{\mathord{\makebox[0pt][r]{$-$}}}
 \providecommand\matlabtextC{\footnotesize\selectfont\strut}%
 \psfrag{000}[ct][ct]{\matlabtextC $0$}%
 \psfrag{001}[ct][ct]{\matlabtextC $20$}%
 \psfrag{002}[ct][ct]{\matlabtextC $40$}%
 \psfrag{003}[ct][ct]{\matlabtextC $60$}%
 \psfrag{004}[ct][ct]{\matlabtextC $80$}%
 \psfrag{005}[ct][ct]{\matlabtextC $100$}%
 \psfrag{013}[ct][ct]{\matlabtextC $\matlabfragNegXTick 30$}%
 \psfrag{014}[ct][ct]{\matlabtextC $\matlabfragNegXTick 20$}%
 \psfrag{015}[ct][ct]{\matlabtextC $\matlabfragNegXTick 10$}%
 \psfrag{016}[ct][ct]{\matlabtextC $0$}%
 \psfrag{017}[ct][ct]{\matlabtextC $10$}%
 \psfrag{018}[ct][ct]{\matlabtextC $20$}%
 \psfrag{019}[ct][ct]{\matlabtextC $30$}%
 %
 %
 %
 \psfrag{006}[rc][rc]{\matlabtextC $4.25$}%
 \psfrag{007}[rc][rc]{\matlabtextC $4.3$}%
 \psfrag{008}[rc][rc]{\matlabtextC $4.35$}%
 \psfrag{009}[rc][rc]{\matlabtextC $4.4$}%
 \psfrag{010}[rc][rc]{\matlabtextC $4.45$}%
 \psfrag{011}[rc][rc]{\matlabtextC $4.5$}%
 \psfrag{012}[rc][rc]{\matlabtextC $4.55$}%
 \psfrag{020}[rc][rc]{\matlabtextC $0$}%
 \psfrag{021}[rc][rc]{\matlabtextC $1$}%
 \psfrag{022}[rc][rc]{\matlabtextC $2$}%
 \psfrag{023}[rc][rc]{\matlabtextC $3$}%
 \psfrag{024}[rc][rc]{\matlabtextC $4$}%
 \psfrag{025}[rc][rc]{\matlabtextC $5$}%
 %
 \includegraphics{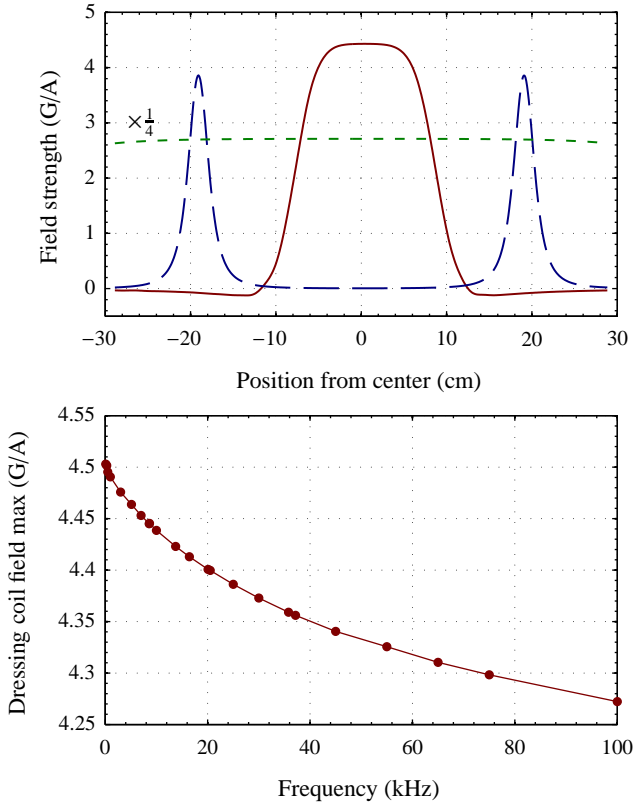}
 \caption{\label{fig:magfields} (Color online) (Top) Calculated magnetic field profiles for the solenoid coil (short-dashed, green curve) and compensating coil (long-dashed, blue curve), along with the experimentally determined profile for the dressing coil in the presence of the magnetic shielding (solid, red curve). (Bottom) The magnitude of the dressing field vs. frequency of the oscillating current in the presence of magnetic shielding.}
\end{figure}

Data were acquired at undressed Larmor precession frequencies of $\omega_0/2\pi = 3.99\kHz$ and $\omega_0/2\pi = 9.36\kHz$.  Figure~\ref{fig:results} shows the compensating field required to maintain minimum transmission as a function of dressing parameter $x$.  The compensating field strength is expressed as its average strength over the entire length of the apparatus divided by the magnetic field required for the undressed Larmor precession frequency, i.e.
\begin{equation}
 \frac{B_{c,av}}{B_0} = \frac{1}{B_0L}\int_{-L/2}^{L/2} B_c(z)\ dz\ ,
\end{equation}
where $B_0 = \omega_0/\gamma$.  For each undressed Larmor precession frequency, multiple sweeps of the intensity of the dressing field with different frequencies were taken, allowing the phase shift to be measured as a function of both dimensionless parameters $x$ and $y$.  Note that the functional form of the prediction mimics the behavior shown in Fig.~\ref{fig:effective_gamma}, but is inverted.

In the region between the coils, we must know the field profiles accurately in order to make a prediction.  Computation of the field profiles due to the solenoid and compensating coils was made using a finite-element solver in the presence of the magnetic shielding.  These calculations were verified using data where, without dressing, the magnitude of the solenoid field was reduced to compensate for the additional field applied by the compensating coils to maintain minimum transmission, paying special attention to the last two integrals in Eq.~\ref{eq:priint}.  The dressing coil field profile was mapped with a pair of orthogonal 7~mm diameter (axial and transverse) search coils and with a Hall probe, both with and without the magnetic shielding, and as a function of frequency.  Not only does the presence of the magnetic shield change the field calibration, but the calibration is dependent upon the frequency of the field generated by the coil.  At these relatively low frequencies, the skin depth is expected to be much larger than the thickness of the surrounding magnetic shield.  However, because of various geometric effects, the penetration is not merely a function of the skin depth alone, resulting in a more complicated dependence on frequency~\cite{Fahy1988,Bidinosti2008}.  The field profiles used in this analysis are shown in Fig.~\ref{fig:magfields}, along with the measured dependence of the dressing coil field on frequency.

In addition to incorporating the influence of the shield on the dressing coil calibration, we include the possibility that the dressing field is not perpendicular to the field of the solenoid and compensating coils.  By applying a static field with the dressing coil rather than an oscillating field, the fields due to the solenoid and the dressing coil add in quadrature.  Let us define the $\hat{x}$ direction as being parallel to the dressing field.  The first integral in Eq.~\ref{eq:priint} then has $\gamma'(z) = \gamma$ (because there is no dressing in this case) and 
\begin{equation}
 B(z) = \sqrt{(B_0\sin\theta + B_c)^2 + (B_{ds} + B_0\cos\theta)^2}\ ,
\end{equation}
where $B_{ds}$ is the static field applied by the dressing coil, $B_0$ is the field applied by the solenoid, $B_c$ is the field due to the compensating coils, and $\theta$ is the angle between the solenoid field and the dressing coil field.   By recording the compensating coil current required to keep the transmission minimum at multiple DC dressing coil currents, we can determine the best fit angle between the dressing coil field and the solenoid field, which is $88.6\pm0.1\deg$.  As shown in Ref.~\cite{Golub1994}, the primary effect of this misalignment when the dressing is active is to only dress the component of the spin precession along the axis perpendicular to the dressing field, while the spin continues to precess normally about the component of the static field parallel to the dressing field.  This effect is included in this analysis.  However, further effects not considered in Ref.~\cite{Golub1994} can result from misalignment and these are discussed in Sec.~\ref{sec:adiabatic}.

For the data shown in Fig.~\ref{fig:results}, we diagonalize the Hamiltonian of Eq.~\ref{eq:ham2} with the misalignment discussed above and determine the effective gyromagnetic ratio by taking the difference of the two central energy eigenstates and dividing by $\hbar B_0$.  This effective gyromagnetic ratio is then determined at each position $z$ in the apparatus.  By assuming that the spin-precession frequency changes quasi-statically, this dressed gyromagnetic ratio appears as a function of $z$, i.e. $\gamma'(z)$ in Eq.~\ref{eq:priint}.  In order to simplify the integrals, we make the assumption that in the region where the correction field $B_c(z)$ is non-negligible [i.e., $B_c(z)/B_0\gg 1$], the precession frequency is set by the undressed $\gamma$, i.e., $\gamma'(z)\approx\gamma$.  We note that this occurs when $B_{c,av}\sim 0.1 B_0$, because $B_c(z)$ has a much smaller spatial extent then $B_0$.  In these regions, the fringe field produced by the dressing coil is $<1$\% of the maximum (dressing) field and so the maximum local value of $x$ incurred is only $x\sim0.08$ when $\omega_0/2\pi = 3.99\kHz$ and $\omega_d/2\pi = 5.12\kHz$.

Using these assumptions, we calculate the required compensating field to maintain minimum transmission.  These theoretical curves are shown in Fig.~\ref{fig:results}.  Note that there are no tunable parameters in the curves shown; all unknown parameters in the theory are determined by calibration.  Discrepancies between the data and the theory are less than 4\% for $1<x<6$ when the undressed Larmor precession frequency is $\omega_0/2\pi = 3.99\kHz$ and less than 2\% for all values of $x$ measured when $\omega_0/2\pi = 9.36\kHz$.  The residuals are not Gaussian and do show some systematic error, which is most likely caused by inaccurate mapping of the magnetic fields.  Any greater precision would require maps of the magnetic fields to be more accurate than 1\%.

\section{Breakdown of the Quasi-Static Approximation}
\label{sec:adiabatic}
The primary assumption made in the analysis presented above is that the spin-precession frequency changes quasi-statically.  One particular region where this assumption can break down is where $\gamma'(z)\rightarrow0$.  In such regions, if the alignment between the dressing field and the static magnetic field is not perfect, an avoided crossing of the eigenvalues of the spin dressing Hamiltonian appears, as illustrated in the inset of Fig.~\ref{fig:states}.  The size of this avoided crossing corresponds to undressed precession of the spin about the static magnetic field that is parallel to the dressing field.

In the sudden, semi-classical approximation used above, the spin precesses about the vector sum of the static field pointing in the same direction as the dressing field ($B_x$, in the $\hat{x}$ direction) with angular frequency $\gamma B_x$ and the component of the static magnetic field orthogonal to the axis of the dressing field ($B_z$, in the $\hat{z}$ direction) with angular frequency $\gamma^\prime(z) B_z$.  Thus, in this approximation, the phase accumulated through the region of interest is
\begin{equation}
 \Delta\phi = \int \sqrt{\left(\frac{d\gamma'(z)}{d z} B_z v t\right)^2 + (\gamma B_x)^2}\ dt\ ,
\end{equation}
where $\gamma$ is the undressed gyromagnetic ratio, $\gamma'(z)$ is the dressed gyromagnetic ratio, $v$ is the velocity, and $t=0$ is the time at which $\gamma'(z)=0$.  Here, the first term represents the precession about $\hat{z}$, where the instantaneous frequency is given by $\omega_z(t) = \gamma'(z(t)) B_z = \frac{d\gamma'(z)}{dz} B_z v t$, and the second term represents the instantaneous precession frequency about $\hat{x}$.  When treating the full precession frequency as merely the (frequency) difference between the energy eigenvalues of the of two states as shown in Fig.~\ref{fig:states}, we implicitly make the assumption that $\omega(-\delta t) = -\omega(\delta t)$, where $\delta t$ is an infinitesimal time step from $t=0$ and $\omega$ is the full angular precession frequency.  Semi-classically, this amounts to time-reversing the spin dynamics regardless of the magnitude of the orthogonal component $B_x$.  Quantum-mechanically, we assume that we are in the strong diabatic transfer limit, because in this picture a state $\ket{\downarrow}$ remains in $\ket{\downarrow}$ regardless of the energy of that state as a function of time.

Rigorous treatment of the effect of $B_x$ complicates the problem significantly for two reasons.  First, because $B_x$ is along $\hat{x}$, the spin will precess outside of the $\hat{x}$-$\hat{y}$ plane in the region where $\gamma'B_z \ll \gamma B_x$.  After traversing this region of interest, the spin can be left in a state with a non-zero value for $\left<S_z\right>$.  Second, if the spin is moving sufficiently slowly through the region of interest, the spin-precession vector will adiabatically follow the vector sum of $\gamma B_x$ and $\gamma' B_z$.  To be specific, at the start of the region of interest, when $\gamma'B_z \gg \gamma B_x$, the spin-precession vector will point along $+\hat{z}$.  As $\gamma'\rightarrow 0$, the spin-precession vector will point along $\hat{x}$, as it will only be precessing about $\gamma B_x$.  As $\gamma'$ becomes negative and its magnitude grows, the spin-precession will realign with the $-\hat{z}$ axis.  The spin therefore accumulates an additional phase of $\pi$ radians as it started along $+\hat{z}$ and ended along $-\hat{z}$.  This additional phase can be approximated by having the spin rotate around a pseudomagnetic field pointing in the $\hat{y}$ direction and appears to be similar to a geometric phase effect.

\begin{figure}
 \center
 %
 %
 \providecommand\matlabtextA{\selectfont\strut}%
 \psfrag{019}[bc][bc]{\matlabtextA $\left<S_z\right>$}%
 \psfrag{020}[bc][bc]{\matlabtextA Phase accumulated (rad)}%
 \psfrag{021}[tc][tc]{\matlabtextA $\gamma B_x \left(\frac{d\gamma'}{dz}B_z v\right)^{-1/2}$}%
 %
 %
 %
 \def\matlabfragNegXTick{\mathord{\makebox[0pt][r]{$-$}}}
 \providecommand\matlabtextB{\footnotesize\selectfont\strut}%
 \psfrag{005}[ct][ct]{\matlabtextB $0$}%
 \psfrag{006}[ct][ct]{\matlabtextB $0.05$}%
 \psfrag{007}[ct][ct]{\matlabtextB $0.1$}%
 \psfrag{008}[ct][ct]{\matlabtextB $0.15$}%
 \psfrag{009}[ct][ct]{\matlabtextB $0.2$}%
 \psfrag{010}[ct][ct]{\matlabtextB $0.25$}%
 \psfrag{011}[ct][ct]{\matlabtextB $0.3$}%
 \psfrag{012}[ct][ct]{\matlabtextB $0.35$}%
 \psfrag{013}[ct][ct]{\matlabtextB $0.4$}%
 %
 %
 %
 \psfrag{000}[rc][rc]{\matlabtextB $-0.6$}%
 \psfrag{001}[rc][rc]{\matlabtextB $-0.4$}%
 \psfrag{002}[rc][rc]{\matlabtextB $-0.2$}%
 \psfrag{003}[rc][rc]{\matlabtextB $0$}%
 \psfrag{004}[rc][rc]{\matlabtextB $0.2$}%
 \psfrag{014}[rc][rc]{\matlabtextB $-1$}%
 \psfrag{015}[rc][rc]{\matlabtextB $-0.5$}%
 \psfrag{016}[rc][rc]{\matlabtextB $0$}%
 \psfrag{017}[rc][rc]{\matlabtextB $0.5$}%
 \psfrag{018}[rc][rc]{\matlabtextB $1$}%
 %
 \includegraphics{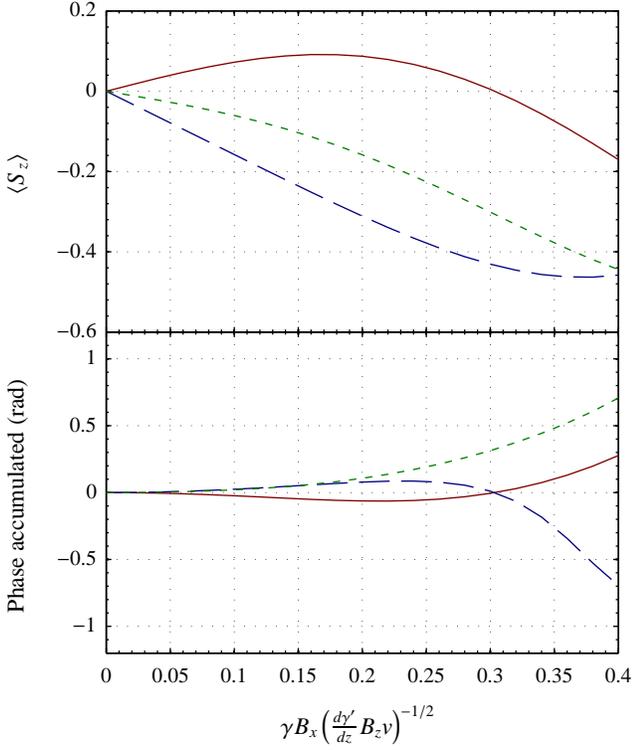}
 \caption{\label{fig:sim1results} (Color online) Results of the simulation of the approximate Hamiltonian (Eq.~\ref{eq:approxdiabaticham}, see text).  In the top pane is the final expectation value for $S_z$ ($-^1/_2 \leq \langle S_z\rangle \leq\text{$^1/_2$}$) and the bottom pane contains the total phase swept out as the spin traverses the avoided crossing.  Note that the simulation is highly dependent on the initial state.  The solid, red lines correspond to an initial state [Eq.~\ref{eq:initialstate}] with $\phi=0$, the blue, dashed lines correspond to $\phi=\pi/2$, and the dotted, green lines correspond to $\phi=\pi/4$.}
\end{figure}

Quantum mechanically, this problem is similar to a Landau-Zener tunneling process, in which the Hamiltonian for the region of interest can be well approximated by
\begin{equation}
\label{eq:approxdiabaticham}
 H = \frac{1}{2} \begin{pmatrix} -\frac{d\gamma'}{dz} B_z v t & \gamma B_x \\ \gamma B_x & \frac{\partial \gamma'}{dz} B_z v t\end{pmatrix}\ .
\end{equation}
In our case, the initial state lies somewhere in the $\hat{x}-\hat{y}$ plane, e.g.
\begin{equation}
 \label{eq:initialstate}
 \ket{\psi} = \frac{1}{\sqrt{2}}\left(e^{-i\phi/2}\ket{\uparrow} + e^{i\phi/2}\ket{\downarrow}\right)\ ,
\end{equation}
where $\phi$ represents the azimuthal angle.  The state is then evolved through the avoided crossing and the additional phase accrued is examined.  To the authors' knowledge, no analytic solution exists to the problem with this initial state.  Nonetheless, we expect that the figure of merit remains the same as for the standard Landau-Zener problem, namely
\begin{equation}
 \gamma^2 B_x^2 \ll \frac{d\gamma'}{dz} B_z v\ ,
\end{equation}
in order for the spin to proceed in the diabatic limit and accumulate no additional phase through the region where $\gamma'\rightarrow0$.  In the present case, $B_z = B_0 \sin\theta$ and $B_x = B_0\cos\theta$, where $\theta = 88.6^\circ$.  Given the angle, we further approximate $B_z\approx B_0$.  The static field $B_0$ can be determined through the undressed precession frequency, e.g. $B_0 = \omega_0/\gamma = f_0/(\gamma/2\pi) \approx 10\kHz/(3.243\mbox{ kHz/G})\approx 3\G$.  As discussed in Sec.~\ref{sec:piby2coils}, the average velocity for the $^3$He was measured to be $v = 140$~m/s.  For the purposes of this estimate, let us consider the case where the amplitude of the dressing field is large enough such that the dressing parameter at $z=0$ (the maximum value of $x(z)$) is given by $x(z=0)\geq2.5$.  Using the amplitude profile shown in Fig.~\ref{fig:magfields}, the $^3$He dressed gyromagnetic ratio will become zero at approximately $z\approx\pm5\cm$.  The rate of change of $\gamma'$ around these points can also be estimated from the amplitude profile in Fig.~\ref{fig:magfields}.  With $d(\gamma'/2\pi)/dz\approx 25\mbox{ (kHz/G)/m}$,
\begin{equation}
 \label{eq:adcond}
 \left(\frac{d\gamma'}{dz}\right)^{-1}\frac{\gamma^2 B_0 \cos^2\theta}{v} \approx 3.5\times 10^{-2}\ ;
\end{equation}
so we are indeed in the diabatic limit.

Simulations were made of the full time-dependent Schr\"{o}dinger equation with the Hamiltonian of Eq.~\ref{eq:approxdiabaticham} to quantify the effect of this misalignment.  The initial state (Eq.~\ref{eq:initialstate}) is propagated from the dimensionless time "$t' = (d\gamma'/dz)B_z v)^{1/2} = -100$ to $t' = 100$, and the total accumulated phase and final values of $\left<S_z\right>$ are computed as a function of $\gamma B_x$.  These results are shown in Fig.~\ref{fig:sim1results} for various initial states.  The final state depends strongly on the initial state.  However, for the diabatic case, we empirically find from simulation that an order of magnitude approximation of the effect is given by
\begin{equation}
 \Delta \phi \approx \pi  \left(\frac{d\gamma'}{dz}\right)^{-1}\frac{\gamma^2 B_0 \cos^2\theta}{v}\ .
\end{equation}
which implies an additional shift of approximately 0.11~rad as the spin traverses the region of interest when $x(z=0)\geq2.5$.  This phase shift is less than 0.1\% of the total phase $\phi$  accumulated by the spin as it traverses between the spin flip coils with no dressing applied. 

Even when the dressing field is perfectly orthogonal to the static field, an additional phase shift to the quasi-static approximation can arise.  As the spin enters the dressing coil, it is subject to a changing magnetic field due to the changing intensity profile of the dressing field (see Fig.~\ref{fig:magfields}).  This effect is well known in interferometer experiments, and a similar effect was discovered by Millman~\cite{Millman1939}.  In the rest frame of the atom, the $^3$He experiences an amplitude-modulated dressing field that changes the dressing field's effective spectral profile causing dressing at multiple frequencies, an effect that is not included in the quasi-static approximation.  In order for the influence of this modulation to be small, the amplitude of the dressing field must change more slowly than the overall precession, so that the spin accurately follows the field as required by the quasi-static approximation.  Simulation of the full Schr\"{o}dinger equation for our field profiles indicates that for $x<3$, this Millman-type effect scales roughly as
\begin{equation}
 \label{eq:condeffect2}
  \Delta\phi \propto \left(\frac{\omega_0}{\omega_d^2}\right) \frac{dx}{dz} v\ .
\end{equation}
Note that this effect will become stronger as the velocity is increased, in contrast to the situation for the avoided crossing effect discussed above.  For this reason, we expect that this Millman-type effect will be the predominant deviation from the quasi-static behavior encountered in our experiment.  One of the largest phase shifts predicted for this experiment is $\Delta\phi \approx -0.8$~rad at $\omega_0/2\pi=3.99\kHz$, $\omega_d/2\pi = 5.12\kHz$, $x(z=0) = 3$, and $v=140\mps$.  This phase shift is slightly less than 1\% of the total phase $\phi$, on the threshold of what might be detectable in our data.

Searching for these effects in our experimental data requires a significant computational effort.  Because these are purely dynamical effects and no analytical solution exists, numerical integration of the Schr\"odinger equation must be made which presents several complications.  First, the field profiles shown in Fig.~\ref{fig:magfields} must be used.  Second, because there is a distribution of velocities and each of the above effects depends strongly on velocity, each velocity class must be simulated separately and subsequently averaged.  Third, when $y\sim 1$, the phase of the spin dressing field will affect the final state; therefore averaging over the phase of the dressing field is also required.  We ran simulations of our Ramsey experiment which calculated the transmitted $^3$He intensity as a function of velocity, the dressing field phase, and $x$ for a given value of $y$.  We first averaged over the phase of the dressing field, assuming that all possible phases are equally likely for each velocity class.  We then averaged over the velocity classes using a Maxwell-Boltzmann distribution with $\left<v\right> = 140\mps$, which corresponds to a temperature of approximately 2.9~K.   For each value of $x$, the simulation tuned the strength of the compensating field to minimize the total transmission.
 
\begin{figure}
 %
 %
 \providecommand\matlabtextA{\selectfont\strut}%
 \psfrag{016}[bc][bc]{\matlabtextA Residuals}%
 \psfrag{017}[tc][tc]{\matlabtextA $x = \gamma B_d/\omega_d$}%
 \psfrag{018}[bc][bc]{\matlabtextA Compensating field ($B_{c,av}/B_0$)}%
 %
 %
 %
 \def\matlabfragNegXTick{\mathord{\makebox[0pt][r]{$-$}}}
 \providecommand\matlabtextB{\footnotesize\selectfont\strut}%
 \psfrag{000}[ct][ct]{\matlabtextB $0$}%
 \psfrag{001}[ct][ct]{\matlabtextB $1$}%
 \psfrag{002}[ct][ct]{\matlabtextB $2$}%
 \psfrag{003}[ct][ct]{\matlabtextB $3$}%
 \psfrag{004}[ct][ct]{\matlabtextB $4$}%
 %
 %
 %
 \psfrag{005}[rc][rc]{\matlabtextB $-4$}%
 \psfrag{006}[rc][rc]{\matlabtextB $-2$}%
 \psfrag{007}[rc][rc]{\matlabtextB $0$}%
 \psfrag{008}[rc][rc]{\matlabtextB $2$}%
 \psfrag{009}[rc][rc]{\matlabtextB $4\times10^{-3}$}%
 \psfrag{010}[rc][rc]{\matlabtextB $0$}%
 \psfrag{011}[rc][rc]{\matlabtextB $0.1$}%
 \psfrag{012}[rc][rc]{\matlabtextB $0.2$}%
 \psfrag{013}[rc][rc]{\matlabtextB $0.3$}%
 \psfrag{014}[rc][rc]{\matlabtextB $0.4$}%
 \psfrag{015}[rc][rc]{\matlabtextB $0.5$}%
 %
 \includegraphics{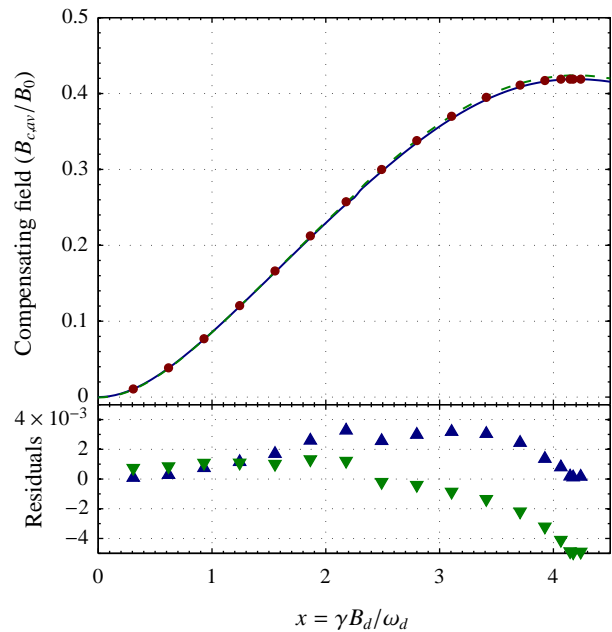}
 \caption{\label{fig:fullsimulation} (Color online) Results of a full simulation integrating the Schr\"{o}dinger equation for $\omega_0/2\pi = 9.36\kHz$ and $\omega_d/2\pi = 13.75\kHz$. Top panel: Data, shown as points, and comparison theory.  The solid blue line shows the quasi-static approximation while the dashed green line shows the full simulation of the experiment. Bottom panel: Difference between data and theory for the quasi-static approximation (blue, upward triangles) and the full simulation (green, downward triangles).}
\end{figure}

Typical results of a simulation are shown in Fig.~\ref{fig:fullsimulation}.  In general, the full quantum simulation better replicates the data for $x<2.3$, but deviates from the data for larger $x$ values.  This deviation appears to be due to the introduction of the avoided crossing effect, which causes rapid oscillations in the transmission vs. velocity curve.  These oscillations could easily lead to numerical errors on the 1\% level.  The additional phase shift that is accumulated in these simulations compared to the quasi-static approximation is due primarily to the Millman-type effect, as the additional phase shift roughly follows Eq.~\ref{eq:condeffect2}.  A comparison of the sum of the squared residuals for the two predictions is inconclusive, as no theory produces a better curve for all data sets.  However, there may be indications that the Millman effect is still seen in the data.  For $x<2.3$ and $y\sim1$, the residuals for the full simulation show little or no $x$ dependence whereas the quasi-static approximation tends to show a linear dependence of the residuals on $x$.  Moreover, the sum of the squared residuals for $x<2.3$ tends to be a factor of 3-4 smaller than that of the quasi-static case.

While easily distinguishable, the absolute difference between the two theoretical results is typically less than 2\%.  In order to conclusively demonstrate these effects experimentally, calibration errors must be suppressed to within 2\%.  Several possible systematics, such as external magnetic fields leaking through the magnetic shielding, improper determination of the magnetic field profiles, and other such effects can easily account for these discrepancies.  However, absolute determination of these systematics is beyond the scope of this work, and therefore conclusions regarding these deviations from the quasi-static condition cannot be made.

\section{Conclusions}

In this work, we verified the spin-dressing Hamiltonian (Eq.~\ref{eq:ham2}) to a level of 2\% in the region of interest for a proposed neutron EDM experiment, verifying the perturbative corrections contained in Ref.~\cite{Golub1994} to the same level.  We also performed simulations to search for breakdown of the quasi-static approximation used above and in~\cite{Golub1994}.  Ultimately, further experimental work (in which systematics are controlled to an absolute level of 1\% or better) would be required to definitively identify effects associated with the breakdown of this approximation.  Alternatively, it could be studied by controlling the atomic velocity distribution. In particular, faster atoms are expected to enhance the magnitude of departures from quasi-static behavior. This approach is not feasible using our present apparatus; in effect we are unable to polarize atoms traveling much faster than those used in this work because of limitations associated with the magnetic fields in the polarizer and analyzer assemblies. One might be able to study the breakdown 
of the quasi-static approximation using a cell-type experiment, such as that presented in Ref.~\cite{Chu2011}. In this case one would use a rapidly modulated dressing field to explore appropriate regions of parameter space.

An understanding of the implications of the breakdown of the quasi-static approximation is required for the proposed neutron EDM experiment~\cite{Golub1994}.  In order to study the effect on the proposed experiment, we simulated the effect of modulation of the spin dressing using realistic parameters.  In the simulation, an undressed Larmor precession rate of 20~Hz was dressed with a modulated dressing parameter, $x = x_c + x_m\cos(\omega_m t)$, where $x_c = 1.31$, $x_m = 0.1$, and $\omega_m/2\pi = 1\Hz$.  The dressing frequency was set to $2\kHz$.  Note that this places us in the regime where $y\ll1$, which greatly suppresses the Millman-type effect discussed in Sec.~\ref{sec:adiabatic}.  Also, because the region near $x=2.3$ is avoided, any misalignment effects are also suppressed.  Because of these suppressions, the quasi-static approximation works quite well to within the numerical errors of the simulation.  No phase lags or unexpected harmonics of the modulation frequency were seen in the simulation.  However, more simulations or experimental work must be done for the final parameters used in the experiment.  This would include simulating a square wave modulation, which would provide a larger signal to noise ratio for the proposed neutron EDM experiment.

\begin{acknowledgments}
The authors would like to thank R. Golub and A.O. Sushkov for useful discussions.  This work was supported by Los Alamos Laboratory Directed Research and Development (2001-2006), and Yale University (2006-2011).  M.E. Hayden was supported by the Natural Sciences and Engineering Research Council of Canada.
\end{acknowledgments}


\begin{thebibliography}{10}%
\makeatletter
\providecommand \@ifxundefined [1]{%
 \ifx #1\undefined \expandafter \@firstoftwo
 \else \expandafter \@secondoftwo
\fi
}%
\providecommand \@ifnum [1]{%
 \ifnum #1\expandafter \@firstoftwo
 \else \expandafter \@secondoftwo
\fi
}%
\providecommand \enquote [1]{``#1''}%
\providecommand \bibnamefont  [1]{#1}%
\providecommand \bibfnamefont [1]{#1}%
\providecommand \citenamefont [1]{#1}%
\providecommand\href[0]{\@sanitize\@href}%
\providecommand\@href[1]{\endgroup\@@startlink{#1}\endgroup\@@href}%
\providecommand\@@href[1]{#1\@@endlink}%
\providecommand \@sanitize [0]{\begingroup\catcode`\&12\catcode`\#12\relax}%
\@ifxundefined \pdfoutput {\@firstoftwo}{%
 \@ifnum{\z@=\pdfoutput}{\@firstoftwo}{\@secondoftwo}%
}{%
 \providecommand\@@startlink[1]{\leavevmode}%
 \providecommand\@@endlink[0]{}%
}{%
 \providecommand\@@startlink[1]{%
  \leavevmode
  \pdfstartlink
   attr{/Border[0 0 1 ]/H/I/C[0 1 1]}%
   user{/Subtype/Link/A<</Type/Action/S/URI/URI(#1)>>}%
  \relax
 }%
 \providecommand\@@endlink[0]{\pdfendlink}%
}%
\providecommand \url  [0]{\begingroup\@sanitize \@url }%
\providecommand \@url [1]{\endgroup\@href {#1}{\urlprefix}}%
\providecommand \urlprefix [0]{URL }%
\providecommand \Eprint[0]{\href }%
\@ifxundefined \urlstyle {%
  \providecommand \doi [1]{doi:\discretionary{}{}{}#1}%
}{%
  \providecommand \doi [0]{doi:\discretionary{}{}{}\begingroup
  \urlstyle{rm}\Url }%
}%
\providecommand \doibase [0]{http://dx.doi.org/}%
\providecommand \Doi[1]{\href{\doibase#1}}%
\providecommand \bibAnnote [3]{%
  \BibitemShut{#1}%
  \begin{quotation}\noindent
    \textsc{Key:}\ #2\\\textsc{Annotation:}\ #3%
  \end{quotation}%
}%
\providecommand \bibAnnoteFile [2]{%
  \IfFileExists{#2}{\bibAnnote {#1} {#2} {\input{#2}}}{}%
}%
\providecommand \typeout [0]{\immediate \write \m@ne }%
\providecommand \selectlanguage [0]{\@gobble}%
\providecommand \bibinfo [0]{\@secondoftwo}%
\providecommand \bibfield [0]{\@secondoftwo}%
\providecommand \translation [1]{[#1]}%
\providecommand \BibitemOpen[0]{}%
\providecommand \bibitemStop [0]{}%
\providecommand \bibitemNoStop [0]{.\EOS\space}%
\providecommand \EOS [0]{\spacefactor3000\relax}%
\providecommand \BibitemShut [1]{\csname bibitem#1\endcsname}%
\bibitem{KhriplovichLamoreaux1997}%
  \BibitemOpen
  \bibfield{author}{%
  \bibinfo {author} {\bibfnamefont{I.}~\bibnamefont{Khriplovich}}\ and\
  \bibinfo {author} {\bibfnamefont{S.}~\bibnamefont{Lamoreaux}},\ }%
  \emph{\bibinfo {title} {CP Violation without Strangeness}}\ (\bibinfo
  {publisher} {Springer-Verlag},\ \bibinfo {address} {Berlin},\ \bibinfo {year}
  {1997})%
  \bibAnnoteFile{NoStop}{KhriplovichLamoreaux1997}%
\bibitem{Farrar1994}%
  \BibitemOpen
  \bibfield{author}{%
  \bibinfo {author} {\bibfnamefont{G.~R.}\ \bibnamefont{Farrar}}\ and\ \bibinfo
  {author} {\bibfnamefont{M.~E.}\ \bibnamefont{Shaposhnikov}},\ }%
  \bibfield{journal}{%
  \Doi{10.1103/PhysRevD.50.774}{\bibinfo {journal} {Phys. Rev. D}}\ }%
  \textbf{\bibinfo {volume} {50}},\ \bibinfo {pages} {774} (\bibinfo {year}
  {1994})%
  \bibAnnoteFile{NoStop}{Farrar1994}%
\bibitem{Peccei1977}%
  \BibitemOpen
  \bibfield{author}{%
  \bibinfo {author} {\bibfnamefont{R.~D.}\ \bibnamefont{Peccei}}\ and\ \bibinfo
  {author} {\bibfnamefont{H.~R.}\ \bibnamefont{Quinn}},\ }%
  \bibfield{journal}{%
  \Doi{10.1103/PhysRevLett.38.1440}{\bibinfo {journal} {Phys. Rev. Lett.}}\ }%
  \textbf{\bibinfo {volume} {38}},\ \bibinfo {pages} {1440} (\bibinfo {year}
  {1977})%
  \bibAnnoteFile{NoStop}{Peccei1977}%
\bibitem{Weinberg1978}%
  \BibitemOpen
  \bibfield{author}{%
  \bibinfo {author} {\bibfnamefont{S.}~\bibnamefont{Weinberg}},\ }%
  \bibfield{journal}{%
  \Doi{10.1103/PhysRevLett.40.223}{\bibinfo {journal} {Phys. Rev. Lett.}}\ }%
  \textbf{\bibinfo {volume} {40}},\ \bibinfo {pages} {223} (\bibinfo {year}
  {1978})%
  \bibAnnoteFile{NoStop}{Weinberg1978}%
\bibitem{Wilczek1978}%
  \BibitemOpen
  \bibfield{author}{%
  \bibinfo {author} {\bibfnamefont{F.}~\bibnamefont{Wilczek}},\ }%
  \bibfield{journal}{%
  \Doi{10.1103/PhysRevLett.40.279}{\bibinfo {journal} {Phys. Rev. Lett.}}\ }%
  \textbf{\bibinfo {volume} {40}},\ \bibinfo {pages} {279} (\bibinfo {year}
  {1978})%
  \bibAnnoteFile{NoStop}{Wilczek1978}%
\bibitem{Golub1994}%
  \BibitemOpen
  \bibfield{author}{%
  \bibinfo {author} {\bibfnamefont{R.}~\bibnamefont{Golub}}\ and\ \bibinfo
  {author} {\bibfnamefont{S.~K.}\ \bibnamefont{Lamoreaux}},\ }%
  \bibfield{journal}{%
  \Doi{10.1016/0370-1573(94)90084-1}{\bibinfo {journal} {Phys. Rep.}}\ }%
  \textbf{\bibinfo {volume} {237}},\ \bibinfo {pages} {1} (\bibinfo {year}
  {1994})%
  \bibAnnoteFile{NoStop}{Golub1994}%
\bibitem{Roberts1973}%
  \BibitemOpen
  \bibfield{author}{%
  \bibinfo {author} {\bibfnamefont{H.~A.}\ \bibnamefont{Roberts}}\ and\
  \bibinfo {author} {\bibfnamefont{F.~L.}\ \bibnamefont{Hereford}},\ }%
  \bibfield{journal}{%
  \Doi{10.1103/PhysRevA.7.284}{\bibinfo {journal} {Phys. Rev. A}}\ }%
  \textbf{\bibinfo {volume} {7}},\ \bibinfo {pages} {284} (\bibinfo {year}
  {1973})%
  \bibAnnoteFile{NoStop}{Roberts1973}%
\bibitem{Surko1970}%
  \BibitemOpen
  \bibfield{author}{%
  \bibinfo {author} {\bibfnamefont{C.~M.}\ \bibnamefont{Surko}}, \bibinfo
  {author} {\bibfnamefont{R.~E.}\ \bibnamefont{Packard}}, \bibinfo {author}
  {\bibfnamefont{G.~J.}\ \bibnamefont{Dick}},\ and\ \bibinfo {author}
  {\bibfnamefont{F.}~\bibnamefont{Reif}},\ }%
  \bibfield{journal}{%
  \Doi{10.1103/PhysRevLett.24.657}{\bibinfo {journal} {Phys. Rev. Lett.}}\ }%
  \textbf{\bibinfo {volume} {24}},\ \bibinfo {pages} {657} (\bibinfo {year}
  {1970})%
  \bibAnnoteFile{NoStop}{Surko1970}%
\bibitem{Passell1966}%
  \BibitemOpen
  \bibfield{author}{%
  \bibinfo {author} {\bibfnamefont{L.}~\bibnamefont{Passell}}\ and\ \bibinfo
  {author} {\bibfnamefont{R.~I.}\ \bibnamefont{Schermer}},\ }%
  \bibfield{journal}{%
  \Doi{10.1103/PhysRev.150.146}{\bibinfo {journal} {Phys. Rev.}}\ }%
  \textbf{\bibinfo {volume} {150}},\ \bibinfo {pages} {146} (\bibinfo {year}
  {1966})%
  \bibAnnoteFile{NoStop}{Passell1966}%
\bibitem{CODATA2006}%
  \BibitemOpen
  \bibfield{author}{%
  \bibinfo {author} {\bibfnamefont{P.~J.}\ \bibnamefont{Mohr}}, \bibinfo
  {author} {\bibfnamefont{B.~N.}\ \bibnamefont{Taylor}},\ and\ \bibinfo
  {author} {\bibfnamefont{D.~B.}\ \bibnamefont{Newell}},\ }%
  \bibfield{journal}{%
  \Doi{10.1103/RevModPhys.80.633}{\bibinfo {journal} {Rev. Mod. Phys.}}\ }%
  \textbf{\bibinfo {volume} {80}},\ \bibinfo {pages} {633} (\bibinfo {year}
  {2008})%
  \bibAnnoteFile{NoStop}{CODATA2006}%
\bibitem{Williams1969}%
  \BibitemOpen
  \bibfield{author}{%
  \bibinfo {author} {\bibfnamefont{W.~L.}\ \bibnamefont{Williams}}\ and\
  \bibinfo {author} {\bibfnamefont{V.~W.}\ \bibnamefont{Hughes}},\ }%
  \bibfield{journal}{%
  \Doi{10.1103/PhysRev.185.1251}{\bibinfo {journal} {Phys. Rev.}}\ }%
  \textbf{\bibinfo {volume} {185}},\ \bibinfo {pages} {1251} (\bibinfo {year}
  {1969})%
  \bibAnnoteFile{NoStop}{Williams1969}%
\bibitem{Flowers1993}%
  \BibitemOpen
  \bibfield{author}{%
  \bibinfo {author} {\bibfnamefont{J.~L.}\ \bibnamefont{Flowers}}, \bibinfo
  {author} {\bibfnamefont{B.~W.}\ \bibnamefont{Petley}},\ and\ \bibinfo
  {author} {\bibfnamefont{M.~G.}\ \bibnamefont{Richards}},\ }%
  \bibfield{journal}{%
  \Doi{10.1088/0026-1394/30/2/004}{\bibinfo {journal} {Metrologia}}\ }%
  \textbf{\bibinfo {volume} {30}},\ \bibinfo {pages} {75} (\bibinfo {year}
  {1993})%
  \bibAnnoteFile{NoStop}{Flowers1993}%
\bibitem{Cohen-Tannoudji1969}%
  \BibitemOpen
  \bibfield{author}{%
  \bibinfo {author} {\bibfnamefont{C.}~\bibnamefont{Cohen-Tannoudji}}\ and\
  \bibinfo {author} {\bibfnamefont{S.}~\bibnamefont{Haroche}},\ }%
  \bibfield{journal}{%
  \Doi{10.1051/jphys:01969003002-3015300}{\bibinfo {journal} {J. Phys.
  France}}\ }%
  \textbf{\bibinfo {volume} {30}},\ \bibinfo {pages} {153} (\bibinfo {year}
  {1969})%
  \bibAnnoteFile{NoStop}{Cohen-Tannoudji1969}%
\bibitem{Esler2007}%
  \BibitemOpen
  \bibfield{author}{%
  \bibinfo {author} {\bibfnamefont{A.}~\bibnamefont{Esler}}, \bibinfo {author}
  {\bibfnamefont{J.~C.}\ \bibnamefont{Peng}}, \bibinfo {author}
  {\bibfnamefont{D.}~\bibnamefont{Chandler}}, \bibinfo {author}
  {\bibfnamefont{D.}~\bibnamefont{Howell}}, \bibinfo {author}
  {\bibfnamefont{S.~K.}\ \bibnamefont{Lamoreaux}}, \bibinfo {author}
  {\bibfnamefont{C.~Y.}\ \bibnamefont{Liu}},\ and\ \bibinfo {author}
  {\bibfnamefont{J.~R.}\ \bibnamefont{Torgerson}},\ }%
  \bibfield{journal}{%
  \Doi{10.1103/PhysRevC.76.051302}{\bibinfo {journal} {Phys. Rev. C}}\ }%
  \textbf{\bibinfo {volume} {76}},\ \bibinfo {pages} {051302} (\bibinfo {year}
  {2007})%
  \bibAnnoteFile{NoStop}{Esler2007}%
\bibitem{Chu2011}%
  \BibitemOpen
  \bibfield{author}{%
  \bibinfo {author} {\bibfnamefont{P.~H.}\ \bibnamefont{Chu}}, \bibinfo
  {author} {\bibfnamefont{A.~M.}\ \bibnamefont{Esler}}, \bibinfo {author}
  {\bibfnamefont{J.~C.}\ \bibnamefont{Peng}}, \bibinfo {author}
  {\bibfnamefont{D.~H.}\ \bibnamefont{Beck}}, \bibinfo {author}
  {\bibfnamefont{D.~E.}\ \bibnamefont{Chandler}}, \bibinfo {author}
  {\bibfnamefont{S.}~\bibnamefont{Clayton}}, \bibinfo {author}
  {\bibfnamefont{B.~Z.}\ \bibnamefont{Hu}}, \bibinfo {author}
  {\bibfnamefont{S.~Y.}\ \bibnamefont{Ngan}}, \bibinfo {author}
  {\bibfnamefont{C.~H.}\ \bibnamefont{Sham}}, \bibinfo {author}
  {\bibfnamefont{L.~H.}\ \bibnamefont{So}}, \bibinfo {author}
  {\bibfnamefont{S.}~\bibnamefont{Williamson}},\ and\ \bibinfo {author}
  {\bibfnamefont{J.}~\bibnamefont{Yoder}},\ }%
  \bibfield{journal}{%
  \Doi{10.1103/PhysRevC.84.022501}{\bibinfo {journal} {Phys. Rev. C}}\ }%
  \textbf{\bibinfo {volume} {84}},\ \bibinfo {pages} {022501} (\bibinfo {year}
  {2011})%
  \bibAnnoteFile{NoStop}{Chu2011}%
\bibitem{Bidinosti2005}%
  \BibitemOpen
  \bibfield{author}{%
  \bibinfo {author} {\bibfnamefont{C.~P.}\ \bibnamefont{Bidinosti}}, \bibinfo
  {author} {\bibfnamefont{I.~S.}\ \bibnamefont{Kravchuk}},\ and\ \bibinfo
  {author} {\bibfnamefont{M.~E.}\ \bibnamefont{Hayden}},\ }%
  \bibfield{journal}{%
  \Doi{10.1016/j.jmr.2005.07.003}{\bibinfo {journal} {J. Magn. Reson.}}\ }%
  \textbf{\bibinfo {volume} {177}},\ \bibinfo {pages} {31} (\bibinfo {year}
  {2005})%
  \bibAnnoteFile{NoStop}{Bidinosti2005}%
\bibitem{BioSavart}%
  \BibitemOpen
  \bibfield{author}{%
  \bibinfo {author} {\bibnamefont{M.W. Reynolds}},\ }%
  \bibinfo {title} {computer code BIOTSAVART Magnetic Field Calculator version 4.0.17},\ (\bibinfo {address} {Ripplon Software, Canada; http://www.ripplon.com}),\ \bibinfo {year} {1992}%
  \bibAnnoteFile{NoStop}{BioSavart}%
\bibitem{Fahy1988}%
  \BibitemOpen
  \bibfield{author}{%
  \bibinfo {author} {\bibfnamefont{S.}~\bibnamefont{Fahy}}, \bibinfo {author}
  {\bibfnamefont{C.}~\bibnamefont{Kittel}},\ and\ \bibinfo {author}
  {\bibfnamefont{S.~G.}\ \bibnamefont{Louie}},\ }%
  \bibfield{journal}{%
  \Doi{10.1119/1.15353}{\bibinfo {journal} {Am. J. Phys.}}\ }%
  \textbf{\bibinfo {volume} {56}},\ \bibinfo {pages} {989} (\bibinfo {year}
  {1988})%
  \bibAnnoteFile{NoStop}{Fahy1988}%
\bibitem{Bidinosti2008}%
  \BibitemOpen
  \bibfield{author}{%
  \bibinfo {author} {\bibfnamefont{C.~P.}\ \bibnamefont{Bidinosti}}\ and\
  \bibinfo {author} {\bibfnamefont{M.~E.}\ \bibnamefont{Hayden}},\ }%
  \bibfield{journal}{%
  \Doi{10.1063/1.2998607}{\bibinfo {journal} {Appl. Phys. Lett.}}\ }%
  \textbf{\bibinfo {volume} {93}},\ \bibinfo {pages} {174102} (\bibinfo {year}
  {2008}),\ ISSN \bibinfo {issn} {00036951}%
  \bibAnnoteFile{NoStop}{Bidinosti2008}%
\bibitem{Millman1939}%
  \BibitemOpen
  \bibfield{author}{%
  \bibinfo {author} {\bibfnamefont{S.}~\bibnamefont{Millman}},\ }%
  \bibfield{journal}{%
  \Doi{10.1103/PhysRev.55.628}{\bibinfo {journal} {Phys. Rev.}}\ }%
  \textbf{\bibinfo {volume} {55}},\ \bibinfo {pages} {628} (\bibinfo {year}
  {1939})%
  \bibAnnoteFile{NoStop}{Millman1939}%
\end{thebibliography}
\end{document}